\documentclass[prl,reprint,twocolumn,showpacs,superscriptaddress,floatfix,aps,10pt]{revtex4-2}
\usepackage{amsmath,amsthm,amssymb}
\usepackage{dcolumn,bm,hyperref,braket}
\usepackage{graphicx}
\usepackage{subfigure,verbatim}
\usepackage{xcolor}
\usepackage{soul}
\usepackage{ulem}
\definecolor{LinkColor}{RGB}{46,48,146}
\usepackage{hyperref}
\hypersetup{
colorlinks=true,
citecolor=LinkColor,
linkcolor=LinkColor,
urlcolor=LinkColor
}

\begin{document}
\title{Chirality-Selective Phonon Pumping by Ferroelectric Dynamics}

\author{Dapeng Yao}
\affiliation{RIKEN Center for Emergent Matter Science (CEMS), 2-1 Hirosawa, Wako, Saitama 351-0198, Japan}
\author{Ping Tang}
\email{tang.ping.a2@tohoku.ac.jp}
\affiliation{Institute for Materials Research, Tohoku University, 2-1-1 Katahira, Sendai, 980-8577, Japan}

\begin{abstract}
The discovery of chiral phonons has expanded the conventional view of lattice vibrations as passive heat carriers, opening new opportunities for phononic spintronic devices. However, their realization has been largely limited to chiral crystals or to specific regions of momentum space in certain achiral materials. Here, we propose a generic mechanism for generating propagating chiral phonons in an ordinary dielectric through the precession of the electric polarization in an adjacent ferroelectric. The polarization dynamics transfers its intrinsic angular momentum to the lattice via electrostrictive coupling, thereby pumping chirality-selective phonons whose handedness is dictated by that of the polarization precession. For a typical LiNbO$_3$$|$Y$_3$Al$_5$O$_{12}$ bilayer, we find that the pumping efficiency quantified by an interfacial convertance substantially exceeds those of thermally induced chiral-phonon generation in chiral crystals, owing to the strong electrostrictive coupling in ferroelectrics. Our work establishes ferroelectric dynamics as a versatile electrical source of chiral phonons and provides a general route toward electrically programmable chiral-phononic and spintronic functionalities.
\end{abstract}
\maketitle

Chiral phonons are circularly polarized lattice vibrations that dynamically break improper rotational symmetry~\cite{Juraschek2025}. Depending on whether opposite enantiomers can be related by time-reversal symmetry, they are classified as falsely or truly chiral, respectively. The discovery of chiral phonons has broadened the conventional view of lattice vibrations as passive heat carriers in solids~\cite{Zhang2015,Zhu2018,Grissonnanche2020,Zhang2022,Ueda2023,Ishito2023,Tsunetsugu2023,Bonini2023,Ohe2024,Yang2025,Che2025}. Their ability to carry angular momentum and couple to other degrees of freedom underpins a variety of fundamental phenomena, including phononic magnetism \cite{Juraschek2019,Cheng2020,Geilhufe2021,Baydin2022,Juraschek2022}, dynamical multiferroicity~\cite{Juraschek2017,Dunnett2019,Basini2024}, the Einstein--de Haas effect~\cite{deHaas1915,Zhang2014,Dornes2019}, and ultrafast demagnetization processes~\cite{PhysRevLett.76.4250,Koopmans2010,Tauchert2022}. Their angular momentum, on the other hand, can be transferred to electronic and magnetic degrees of freedom, inducing electronic spin current~\cite{Hamada2020,Kim2023,YaoAPL2024,Ohe2024,Funato2024,Yao2025,Nishimura2025}, charge current~\cite{Yao2022}, orbital magnetization~\cite{Xiao2021,Ren2021,YaoOAM2025}, and magnetic resonance excitation~\cite{Ren2024,Yao2024,Yokoyama2024,Royo2026,Hwang2026}.
However, truly chiral phonons typically occur either in chiral crystals (e.g., quartz) propagating along the screw axis~\cite{Hamada2018,Ishito2023,Ueda2023,Ohe2024,ZhangNP2025}, where their chirality is dictated by the structural handedness, or at specific wave vectors in the Brillouin zone of certain achiral systems~\cite{j7bs-2zbx,ueda2025chiral,gcgl-9sbb,yang2026symmetry}, thereby substantially restricting their practical applications. Moreover, the experimental signals of chiral phonons are often accompanied by electronic contributions~\cite{Kim2023,Ohe2024,Nabei2026,Nuomin2026}, making it challenging to isolate their intrinsic properties. Therefore, an efficient approach to generating pure (truly) chiral phonons in \textit{ordinary} dielectrics is highly desirable, not only for exploring fundamental chirality-related physics but also for developing chiral-phonon-based spintronic functionalities.

Ferroelectrics are characterized by a spontaneous electric polarization that provides a versatile degree of freedom for controlling material states by an external electric field. The recent discoveries of ferrons~\cite{Tang2022,Bauer2022,Bauer2023,Tang2024,choe2026observation,zhang2025electric,shen2025observation,itoh2026observation} and multiferrons~\cite{Tang2026,Pols2026} as collective quasiparticle excitations of ferroelectric order have established ferroelectrics as a promising platform for unconventional thermoelectric transport~\cite{Bauer2021,TangPRL2022,wooten2023electric,PhysRevB.107.L121406,3y1m-66s1,shen2025observation,itoh2026observation,lopez2026ferron} and dynamical phenomena~\cite{PhysRevB.109.134307,zhou2023surface,choe2026observation,zhang2025electric,Tang2026}. In particular, multiferrons associated with the precessional dynamics of ferroelectric polarization carry both an intrinsic electric dipole moment and orbital angular momentum~\cite{Tang2026,Pols2026}. By analogy with spin pumping in magnetic systems~\cite{Tserkovnyak2002,Tserkovnyak2005,Saitoh2006,Kajiwara2010}, where magnetization dynamics injects spin angular momentum into an adjacent nonmagnetic layer, it remains an entirely open question whether a precessing electric polarization can pump chiral phonons by transferring its orbital angular momentum to the lattice.

\begin{figure}
\begin{center}
\includegraphics[width=8cm]{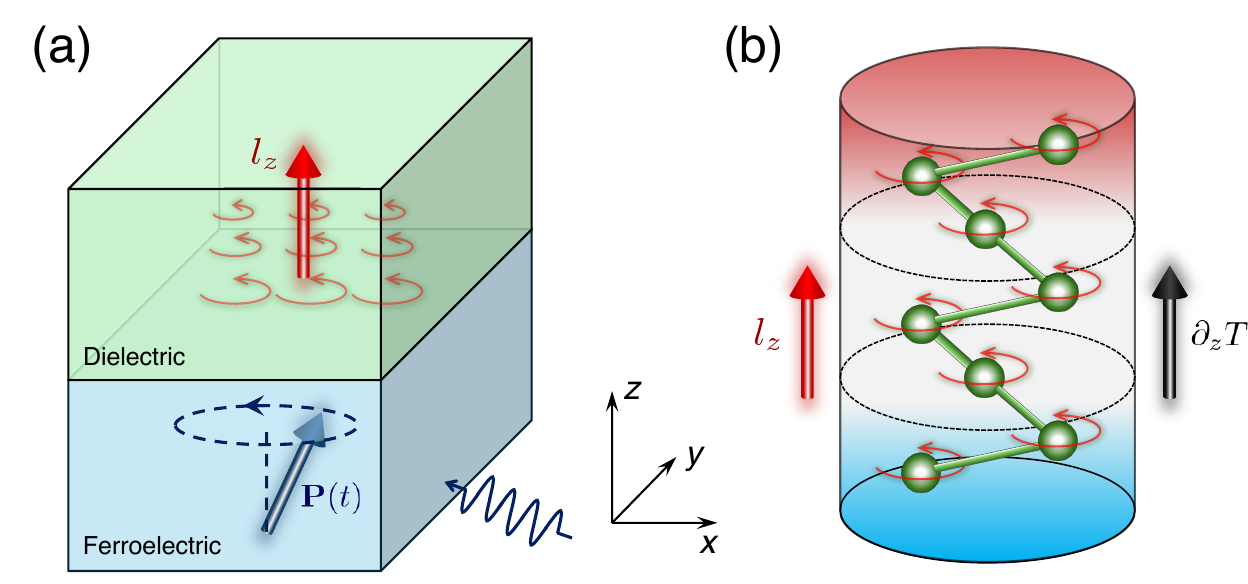}
\caption{(a) Schematic of chirality-selective phonon pumping in a ferroelectric$|$dielectric bilayer proposed in this Letter. The precessing electric polarization $\mathbf{P}(t)$ in the $xy$ plane transfers its angular momentum with a definite chirality to the lattice via electrostrictive coupling, generating chiral phonons that propagate along the $z$ direction in the dielectric with a net angular-momentum density polarized along the $z$ axis. (b) Thermally induced chiral-phonon generation in chiral crystals via the phonon thermal Edelstein effect~\cite{Hamada2018,ZhangNP2025}. A temperature gradient $\partial_z T$ applied along the screw ($z$) axis induces PAM polarized along the $z$ direction. In both cases, the chiral phonons exhibit true chirality, with their angular momentum aligned with the propagation direction.
}
\label{fig1}
\end{center}
\end{figure}

In this Letter, we propose a generic mechanism for generating chirality-selective phonons into an ordinary dielectric by the polarization dynamics in an adjacent ferroelectric layer, as illustrated in Fig.~\ref{fig1}(a). The precessing polarization transfers its intrinsic angular momentum to the lattice through electrostrictive coupling, thereby pumping chiral phonons propagating in the dielectric layer. Based on continuum elastic theory, we formulate the response of the phonon angular momentum (PAM) density to the polarization dynamics in terms of an interfacial angular-momentum convertance. As a representative example, we demonstrate highly efficient chiral-phonon pumping in a LiNbO$_3|$Y$_3$Al$_5$O$_{12}$ (YAG) heterostructure. Owing to the strong electrostrictive coupling in ferroelectrics, the predicted pumping efficiency substantially exceeds that of magnetoelastic phonon pumping~\cite{Sato2021} and thermally induced chiral-phonon generation in chiral crystals~\cite{Hamada2018,ZhangNP2025}~[see Fig.~\ref{fig1}(b)]. Moreover, the chirality of the pumped phonons is controllable through the handedness of the electric polarization precession while it can only be fixed by the structural handedness in chiral crystals. Our work opens a general route toward the generation of \textit{chirality-selective} phonons in ordinary dielectrics, offering new opportunities for exploring the physics of PAM and developing chiral-phononic functionalities.

We consider a bilayer composed of a ferroelectric at $z<0$ with the spontaneous ferroelectric polarization normal to the interface and a dielectric at $z>0$, as shown in Fig.~\ref{fig1}(a). The elastic Lagrangian density of the bilayer system reads
\begin{align}
\mathcal{L}=\frac{1}{2}\rho\dot{\mathbf{u}}^2-\frac{1}{2}C_{ijkl}\varepsilon_{ij}\varepsilon_{kl}+q_{ijkl}\varepsilon_{ij}P_kP_l \left[1-\Theta(z)\right],
\end{align}
where $\rho$ is the mass density, $\varepsilon_{ij}=\frac{1}{2}(\partial u_j/\partial x_{i}+\partial u_i/\partial x_{j})$ is the symmetric strain tensor, with $u_i$ being the $i$th component of the phonon displacement $\mathbf{u}$, and $C_{ijkl}$ is the elastic stiffness tensor. The last term describes the electrostrictive coupling, which is present in the ferroelectric. Here $q_{ijkl}$ is a fourth-rank electrostriction tensor, $P_k$ is the $k$th component of the electric polarization $\mathbf{P}$ in the ferroelectric, and $\Theta(z)$ is the Heaviside step function defined as $\Theta(z>0)=1$ and $\Theta(z<0)=0$. The Einstein summation convention is implied over repeated Cartesian indices. For simplicity, here we assume identical elastic properties (i.e., $\rho$ and $C_{ijkl}$) for the ferroelectric and dielectric layers, although the main conclusion remains valid without this assumption.

From the Euler-Lagrange equation, the equation of motion for the phonon displacements reads
\begin{align}
\rho\ddot{u}_i-C_{ijkl}\frac{\partial \varepsilon_{kl}}{\partial x_{j}}=q_{izkl}P_{k}P_{l}\delta(z), \label{EOM}
\end{align}
where the right-hand-side term comes from the electrostrictive coupling and serves as a phonon-pumping source at the interface. Because of the translational symmetry parallel to the interface, the polarization dynamics excites the phonons that propagates normal to the interface (along the $z$ axis) only. The displacement is then expressed as
\begin{align}\label{gen_solu}
\mathbf{u}(z,t)=\frac{1}{\sqrt{\rho V}}\sum_{q\sigma}Q_{q\sigma}(t)\hat{\mathbf{e}}_{q\sigma}e^{iqz},
\end{align}
where $V$ is the system volume, and $Q_{q\sigma}(t)$ and $\hat{\mathbf{e}}_{q\sigma}$ denote the normal coordinate and polarization vector with the phonon mode $\sigma$ at the wave vector $q\hat{\mathbf{e}}_z$, respectively.
Considering an elastically isotropic medium, the elastic stiffness tensor reduces to $C_{ijkl}=\lambda\delta_{ij}\delta_{kl}+\mu(\delta_{ik}\delta_{jl}+\delta_{il}\delta_{jk})$, with the two Lam\'e parameters $\mu$ and $\lambda$, and the phonon spectrum consists of two degenerate transverse modes with polarization vectors $\hat{\mathbf{e}}_{T_{1}}=\hat{\mathbf{e}}_{x}$ and $\hat{\mathbf{e}}_{T_{2}}=\hat{\mathbf{e}}_{y}$, and one longitudinal mode with $\hat{\mathbf{e}}_{L}=\hat{\mathbf{e}}_{z}$. Substituting Eq~(\ref{gen_solu}) into Eq.~(\ref{EOM}) yields
\begin{align}\label{eq_Q}
\ddot{Q}_{q\sigma}(t)+\omega_{q\sigma}^2Q_{q\sigma}(t)=F_{q\sigma}(t),
\end{align}
where $\omega_{q\sigma}=c_{\sigma} \vert q\vert$ is the acoustic phonon dispersion of branch $\sigma$, with transverse and longitudinal sound velocities $c_{T}=\sqrt{\mu/\rho}$ and $c_{L}=\sqrt{(\lambda+2\mu)/\rho}$, respectively. The corresponding driving force is given by
\begin{align}
F_{q\sigma}(t)=\frac{S}{\sqrt{\rho V}}q_{\sigma zkl} P_k(t)P_l(t),
\end{align}
where $S$ is the interface area, and $\sigma=x,y,z$ labels the phonon branches polarized along the vectors $\hat{\mathbf{e}}_{\sigma}$ in the Cartesian basis. The formal solution of Eq.~(\ref{eq_Q}) is
\begin{align}
Q_{q\sigma}(t)=-\frac{1}{2\omega_{q\sigma}}\int_{0}^{\infty} dt'G_{q\sigma}(t')F_{q\sigma}(t-t'), \label{sol}
\end{align}
where $G_{q\sigma}(t)=-2\Theta(t)e^{-t/2\tau}\sin(\omega_{q\sigma}t)$ is the retarded Green's function of phonons. In the frequency domain, $G_{q\sigma}(\omega)=2\omega_{ q\sigma}/[(\omega+i/2\tau)^2-\omega^2_{q\sigma}]$, where $\tau$ is the phenomenological phonon relaxation time.

We now evaluate the PAM in the dielectric pumped by a polarization dynamics $\mathbf{P}(t)$ from the ferroelectric. Using Eqs.~(\ref{gen_solu}) and (\ref{sol}), the associated PAM density reads
\begin{align}\label{j_ave}
 \bm{l}(z,t)&=\rho \mathbf u(z,t)\times\dot{\mathbf u}(z,t) \nonumber\\
&=\frac{1}{4V}\sum_{qq^{\prime}\sigma\sigma'}\frac{e^{i(q+q')z}}{\omega_{q\sigma}\omega_{q^{\prime}\sigma^{\prime}}}(\hat{\mathbf e}_{\sigma}\times\hat{\mathbf e}_{\sigma'})\int_0^{\infty}dt_1\int_0^{\infty}dt_2\nonumber\\
&\times G_{q\sigma}(t_1)G_{q'\sigma'}(t_2)F_{q\sigma}(t-t_{1})\dot{F}_{q^{\prime}\sigma^{\prime}}(t-t_{2}),
\end{align}
which contains both ac and dc components. In the following, we focus on the dc component, which survives time averaging and represents the intrinsic (``spin") angular momentum of chiral phonons~\cite{Zhang2014}.
Here we consider a monochromatic electric-polarization dynamics $\mathbf{P}(t)=P_{0}\hat{\mathbf{e}}_z+\text{Re}[\mathbf{A}e^{-i\omega t}]$, where $\delta\mathbf{P}(t)=\text{Re}[\mathbf{A}e^{-i\omega t}]$ represents a small fluctuation with frequency $\omega$ and amplitude $\mathbf{A}$. To second order in $\delta\mathbf{P}(t)$, the time-averaged dc component of $\bm l(z,t)$ is given by
\begin{align}
\langle \bm{l}(z)\rangle_{\text{dc}}\simeq \frac{2\omega P_0^2}{\rho}\operatorname{Im}\sum_{\sigma,\sigma^{\prime}}\mathcal{C}_{\sigma\sigma^{\prime}}\Gamma_{\sigma}^{\ast}(z)\Gamma_{\sigma^{\prime}}(z)\hat{\mathbf{e}}_{\sigma}\times\hat{\mathbf{e}}_{\sigma^{\prime}}, \label{dc}
\end{align}
where $\langle\cdots\rangle_{\mathrm{dc}}$ denotes the time average extracting the dc part, $\mathcal{C}_{\sigma\sigma^{\prime}}=q_{\sigma zzl} q_{\sigma^{\prime}zz k} A_{l}^{\ast} A_{k}$, and
\begin{equation}
 \Gamma_{\sigma}(z)=\int \frac{d q}{2\pi}\frac{e^{iqz}}{\omega_{q\sigma}^2-(\omega+i/2\tau)^2}=\frac{ie^{i\tilde{q}_{\sigma}\vert z \vert}}{2c_{\sigma}^2\tilde{q}_{\sigma}},  
\end{equation}
with $\tilde{q}_{\sigma}\equiv (\omega+i/2\tau)/c_{\sigma}$ being a complex-valued wave vector. For concreteness, we consider a uniaxial ferroelectric such as LiNbO$_3$, where the relevant components of the electrostrictive tensor read $q_{zzzz}\equiv g_{\parallel}$, $q_{xzzx}=q_{yzzy}\equiv g_{\perp}$, and $q_{xzzy}=q_{yzzx}=0$~\cite{PhysRevB.71.184110}. The different components of Eq.~(\ref{dc}) then reduce to
\begin{align}
\langle l_{z}(z)\rangle_{\text{dc}}=&\frac{ g_{\perp}^2 P_{0}^2}{\rho c_{T}^2}\frac{\omega e^{-\vert z\vert/\lambda_{T}}}{\omega^2+(1/2\tau)^2}\text{Im}[A_{x}^{\ast}A_{y}], \label{lz}\\
\langle l_{x}(z)\rangle_{\text{dc}}=&\frac{ g_{\parallel}g_{\perp}P_{0}^2 }{\rho c_{L}c_{T}}\frac{\omega e^{-\vert z\vert/\lambda_{\text{av}}}}{\omega^2+(1/2\tau)^2} \vert A_{y}^{\ast}A_{z}\vert \sin (\phi_{x} -q_c \vert z\vert), \label{lx}\\
\langle l_{y}(z)\rangle_{\text{dc}}=&\frac{ g_{\parallel}g_{\perp}P_{0}^2}{\rho c_{L}c_{T}}\frac{\omega e^{-\vert z\vert/\lambda_{\text{av}}}}{\omega^2+(1/2\tau)^2} \vert A_{z}^{\ast}A_{x}\vert  \sin(\phi_{y}+ q_c \vert z\vert), \label{ly}
\end{align}
where $\lambda_{\sigma}=c_{\sigma}\tau$ is the phonon mean-free path of mode $\sigma$, $\lambda_{\text{av}}=2\lambda_{T}\lambda_{L}/(\lambda_{T}+\lambda_{L})$ is the effective mean-free path for longitudinal-transverse hybrid modes, $q_{c}=\omega (c_{T}^{-1}-c_{L}^{-1})$ is a characteristic wave number, and $\phi_{x}=\text{arg} [A_{y}^{\ast}A_{z}]$ and $\phi_{y}=\text{arg} [A_{z}^{\ast}A_{x}]$ are corresponding relative phases. The longitudinal component in Eq.~(\ref{lz}) is carried by the superposition of two degenerate transverse phonon modes and decays exponentially with propagation distance. 
By contrast, the transverse components in Eqs.~(\ref{lx}) and~(\ref{ly}) exhibit an oscillatory decay originating from acoustic birefringence between the longitudinal and transverse phonon modes, characterized by the wave vector $q_{c}$.

When the polarization dynamics is confined to a plane parallel to the interface, only the longitudinal component is generated and corresponds to truly chiral phonons formed by circularly polarized transverse modes that carry angular momentum along their propagating direction. Analogous to spin pumping by magnetization dynamics~\cite{Tserkovnyak2002,Tserkovnyak2005}, the pumped PAM density at the interface can be rewritten as
\begin{equation}
\langle \bm l(z=0)\rangle_{\text{dc}}= \chi_{\text{ph}}(\omega)\delta\mathbf{P}(t)\times \delta\dot{\mathbf{P}}(t), \label{chiral}
\end{equation}
where $\delta \mathbf{P}\times \delta\dot{\mathbf{P}}=\omega \text{Im} (A_{x}^{\ast}A_{y})$ corresponds to the intrinsic angular momentum of the polarization dynamics~\cite{Tang2026}, and $\chi_{\text{ph}}(\omega)=g_{\perp}^2P_{0}^2/[\rho c_{T}^2(\omega^2+(1/2\tau)^2)]$ defines an interfacial angular-momentum convertance that quantifies the efficiency of angular-momentum transfer from the polarization dynamics to the lattice. The in-plane polarization precession dynamics can be excited electrically using interdigital transducers (IDTs), similar to the piezoelectric generation of surface acoustic waves~\cite{White1965}. Unlike the surface acoustic waves, the pumped phonons are truly chiral~\cite{Juraschek2025}, with their angular momentum aligned with the wave vector, as characteristic of chiral phonons in chiral crystals~\cite{Ishito2023,Ohe2024}; moreover, their chirality is selected by the handedness of the polarization precession, i.e., the sign of the precession frequency $\omega$. The response in Eq.~(\ref{chiral}) closely resembles phononic spin pumping by magnetization precession~\cite{PhysRevB.106.014407,PhysRevLett.121.027202}, with the polarization dynamics acting as a source of angular momentum and the lattice as its recipient. Since electrostrictive coupling in ferroelectrics is typically much stronger than magnetoelastic coupling in solids, we shall show the polarization dynamics can provide a highly efficient route for generating chiral phonons.

\begin{figure}
\begin{center}
\includegraphics[width=8.5cm]{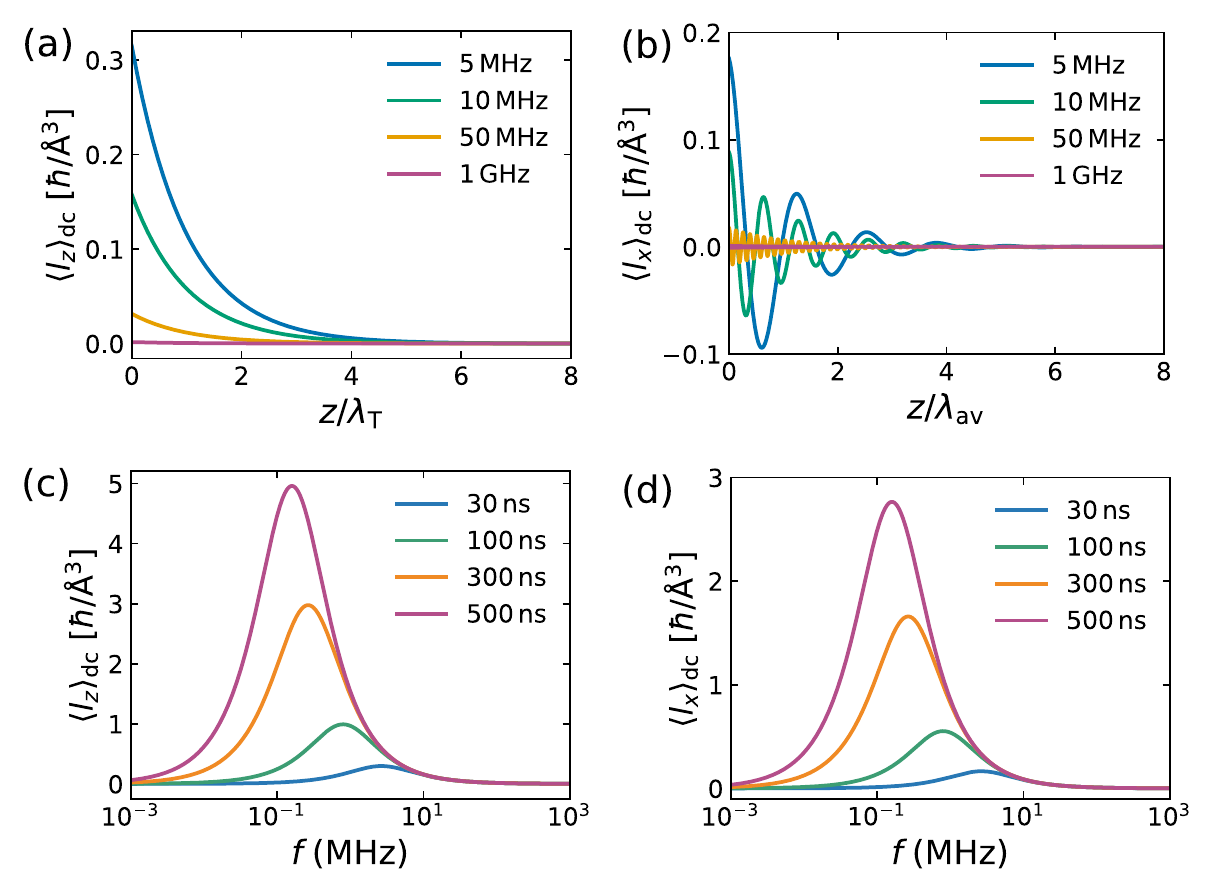}
\caption{Spatial distribution and frequency dependence of the induced PAM density in the YAG layer. (a) Longitudinal component $\langle l_z(z)\rangle_{\mathrm{dc}}$ as a function of $z/\lambda_T$, induced by the in-plane polarization precession $\delta\mathbf{P}(t)$ shown in Fig.~\ref{fig1}, and (b) transverse component $\langle l_x(z)\rangle_{\mathrm{dc}}$ as a function of $z/\lambda_{\mathrm{av}}$, induced by the polarization precession $\delta\mathbf{P}^{\prime}(t)$ in the $yz$ plane, at different driving frequencies. The phonon relaxation time is fixed at $\tau=300\,\mathrm{ns}$. Frequency dependence of (c) the longitudinal component $\langle l_z\rangle_{\mathrm{dc}}$ and (d) the transverse component $\langle l_x\rangle_{\mathrm{dc}}$ at the interface for different phonon relaxation times. 
}
\label{fig2}
\end{center}
\end{figure}

Figures~\ref{fig2}(a) and \ref{fig2}(b) present the spatial distribution of the induced PAM density over the frequency range from $5$ MHz to $1$ GHz in a YAG dielectric layer in contact with LiNbO$_3$, where the two materials have comparable elastic properties. The parameters used in the calculations are $g_{\parallel}=1.848\times10^{9}$ N m$^2$/C$^2$ and $g_{\perp}=1.95\times10^9$ N m$^2$/C$^2$ for LiNbO$_3$~\cite{PhysRevB.71.184110}, and $\rho=4.6\times10^3$ kg/m$^3$~\cite{Weis1985}, $c_L=8.5\times10^3$ m/s, and $c_T=5.0\times10^3$ m/s for YAG~\cite{Balodhi2020}. 
Given the low acoustic attenuation of YAG in the GHz regime~\cite{Dutoit1974}, we adopt a constant relaxation time $\tau=300\,\mathrm{ns}$, corresponding to a transverse phonon mean free path $\lambda_T=1.5\,\mathrm{mm}$ and an effective mean free path $\lambda_{\mathrm{av}}=1.89\,\mathrm{mm}$. The spontaneous polarization of LiNbO$_3$ is taken as $P_0=0.75$ C/m$^2$~\cite{Gopalan1998,Sanna_2017}. We first introduce an in-plane circular polarization precession, $\delta\mathbf P(t)=A_0(\cos\omega t\hat{\mathbf{e}}_x+\sin\omega t\hat{\mathbf{e}}_y)$, with $A_0=7.5\times10^{-3}$ C/m$^2$ being 1\% of the spontaneous polarization in the LiNbO$_3$ layer. As shown in Fig.~\ref{fig2}(a), the resulting longitudinal component $\langle l_z(z)\rangle$ decays exponentially away from the interface. For comparison, the transverse component $\langle l_x(z)\rangle$ generated by an out-of-plane circular polarization precession in the $yz$ plane, $\delta\mathbf{P}'(t)=A_0(\cos\omega t\hat{\mathbf{e}}_y+\sin\omega t\hat{\mathbf{e}}_z)$, are shown in Fig.~\ref{fig2}(b), which exhibits an oscillatory decay with a spatial period $2\pi/|q_c|$.~In both cases, the induced PAM density reaches the order of $0.1\hbar/\mathrm{\AA}^{3}$ near the interface, substantially exceeding that reported in Ref.~\cite{Tang2026,Hamada2018,ZhangNP2025}. Accordingly, the frequency dependences of the longitudinal and transverse components at the interface ($z=0$) for different relaxation times are shown in Figs.~\ref{fig2}(c) and \ref{fig2}(d), respectively. Both components exhibit a characteristic maximum at the optimal (linear) frequency $f_{\text{op}}=\omega_{\mathrm{op}}/2\pi=1/(4\pi\tau)$, reflecting a balance between the increasing angular-momentum injection with driving frequency and its limitation by phonon relaxation. 

We compare the efficiency of chiral-phonon generation by ferroelectric dynamics with that induced by a temperature gradient in chiral crystals. In thermal equilibrium, the angular momenta of phonons at different wave vectors in chiral crystals cancel upon summation over momentum space due to time-reversal symmetry~\cite{Hamada2018,Ohe2024,ZhangNP2025}. As shown in Fig.~\ref{fig1}(b), a temperature gradient ($\partial_z T$) applied along the screw ($z$) axis drives the phonon distribution out of equilibrium, leading to a net PAM density $\langle l_z\rangle=\alpha_{zz}\partial_z T$ through the so-called phonon thermal Edelstein effect~\cite{Hamada2018,ZhangNP2025}. For chiral tellurium (Te), the response coefficient $\alpha_{zz}$, which quantifies the efficiency of thermal PAM generation, is estimated to be $\alpha_{zz}=4.8\times10^{-7}\times[\tau_{\text{Te}}/\mathrm{s}]\,$J\,s\,m$^{-2}$ K$^{-1}$~\cite{Hamada2018}, where $\tau_{\text{Te}}$ is the phonon relaxation time in Te~\cite{Cooper1969,Gerlach1979,Hamada2018,ZhangNP2025}. For a temperature gradient of $\partial_z T=60$ K/mm, as employed in a recent experiment~\cite{ZhangNP2025}, the resulting PAM density is $\langle l_z^{\text{Te}}\rangle\simeq2.88\times10^{-2}\times[\tau_{\text{Te}}/\mathrm{s}]$ J s/m$^3$. In contrast, Eq.~(\ref{chiral}) gives the PAM density generated by ferroelectric dynamics at the optimal frequency, $\langle l_z^{\text{FE}}\rangle=\omega_{\text{op}}A_{0}^2\chi_{\text{ph}}(\omega_{\text{op}})\simeq10^3\times[\tau/\mathrm{s}]\,$J s/m$^3$. Assuming comparable phonon relaxation times, we obtain $\langle l_z^{\text{FE}}\rangle/\langle l_z^{\text{Te}}\rangle\simeq3.5\times10^{4}$, suggesting that chiral-phonon pumping by ferroelectric dynamics can be approximately \textit{four} orders of magnitude larger than that by thermal gradient in chiral crystals. Moreover, in contrast to the phonon thermal Edelstein effect~\cite{Hamada2018,ZhangNP2025}, the present mechanism enables the pumping of chiral phonons with a prescribed chirality set by the polarization precession. These results demonstrate that ferroelectric dynamics provides an efficient and electrically tunable means of generating propagating phonon angular momentum in dielectrics without intrinsic structural chirality.

In conclusion, we propose an efficient electrical mechanism for generating pure  chiral phonons propagating in \textit{ordinary} dielectrics through the ferroelectric polarization dynamics. The mechanism relies on the transfer of angular momentum from the polarization precession to the lattice via the ubiquitous electrostrictive coupling in ferroelectric materials, establishing the ferroelectric counterpart of spin pumping by magnetization dynamics in spintronics. For a typical LiNbO$_3$$|$YAG bilayer, we find that the pumping efficiency quantified in terms of an interfacial convertance substantially exceeds thermally induced chiral-phonon generation in chiral crystals~\cite{Hamada2018,ZhangNP2025} due to the generally strong electrostriction in ferroelectrics. 

Unlike existing chiral phonons, which arise along the screw axis in chiral crystals or at specific wave vectors in achiral systems, the generated chiral phonons in our proposal do not rely on structural chirality and can span a broad frequency range. Instead, their chirality is selected by the handedness of the polarization precession and can therefore be switched electrically by reversing the precession direction, providing a generic and electrically controllable source of chiral phonons in ordinary dielectrics. The pumped chiral phonons propagate with a macroscopic net angular momentum density that may be detected electrically through the inverse spin Hall effect in an adjacent heavy metal~\cite{Kim2023,Ohe2024,Funato2024} or optically by circular dichroism~\cite{Zhu2018,Ueda2023}. Beyond providing a versatile platform for exploring the fundamental physics of chiral phonons, our work opens new opportunities for chiral-phonon-based spintronic functionalities, including all-electrical generation of chiral phonons and phonon-mediated magnetization manipulation or switching in magnetic materials.

\begin{acknowledgments}
\textit{Acknowledgments---}
D.Y. was supported by Japan Society for the Promotion of Science (JSPS) KAKENHI Grant No.~JP25K23366 and RIKEN Special Postdoctoral Researchers Program. P.T. was supported by JSPS KAKENHI Grant-in-Aid for Scientific Research (B) Grant No.~26K00625.
\end{acknowledgments}

%apsrev4-2.bst 2019-01-14 (MD) hand-edited version of apsrev4-1.bst
%Control: key (0)
%Control: author (8) initials jnrlst
%Control: editor formatted (1) identically to author
%Control: production of article title (0) allowed
%Control: page (0) single
%Control: year (1) truncated
%Control: production of eprint (0) enabled
%

%\bibliography{ferroelectric.bib}

\end{document}